\documentclass[aip,
reprint,%
]{revtex4-1}

\usepackage{graphicx}
\usepackage{dcolumn}
\usepackage{bm}
\begin{document}

\title{Dual top gated graphene transistor in the quantum Hall regime}

\author{Ajay K. Bhat}
\affiliation{Department of Condensed Matter Physics and Materials
Science, Tata Institute of Fundamental Research, Homi Bhabha Road,
Mumbai, 400005 India}
\affiliation{Birla Institute of Technology and Science, Vidya Vihar, Pilani, 333031 India}
\author{Vibhor Singh}
\email{vibhor@tifr.res.in}
\author{Sunil Patil}
\author{Mandar~M.~Deshmukh}
\affiliation{Department of Condensed Matter Physics and Materials
Science, Tata Institute of Fundamental Research, Homi Bhabha Road,
Mumbai, 400005 India}
\begin{abstract}
We study the effect of local modulation of charge density and carrier type in graphene field effect transistors using a double top gate geometry. The two top gates lead to the formation of multiple \emph{p-n} junctions. Electron transport measurements at low temperature and in the presence of magnetic field show various integer and fractionally quantized conductance plateaus. We explain these results based on the mixing of the edge channels and find that inhomogeneity plays an important role in defining the exact quantization of these plateaus, an issue critical for the metrology applications of \emph{p-n} junctions.
\end{abstract}


\maketitle

The electronic properties of graphene have been studied extensively in recent years \cite{CastronetoRMP}, including effects like anomalous integer quantum Hall effect \cite{novoselov_two-dimensional_2005,zhang_experimental_2005}. The advantage of controlling the charge type and electric field locally adds a new dimension to study electron transport \cite{young_electronic_2011} to see effects like Klein tunneling \cite{Young_interference,Huard_Klein}, Andreev reflection \cite{BeenakkerRMP}, collapse of Landau levels \cite{NunGuCollapseLL}, Veselago lens \cite{cheianov_focusing_2007} and collimation of electrons with top gates \cite{MarcusCollimation}. The top gate geometry has been utilized in controlling the edge channels in the quantum Hall regime and with control over local and global carrier density. Such \emph{p-n} \cite{MarcusPN,Huard_Klein} and \emph{p-n-p} junctions \cite{oezyilmaz_electronic_2007} show integer and fractional quantized conductance plateaus. These integer and fractional quantized plateaus have been explained with the reflection and mixing of the edge channels leading to the partition of the current \cite{abanin_quantized_2007}.

In this letter, we study electron transport in a graphene multiple lateral heterojunction device with charge density distribution of the type \emph{q-q$_1$-q-q$_2$-q} with independent and complete
control over both the charge carrier type and density in the three different regions. This is achieved by using a global back gate (BG) to fix the overall carrier type and density and local top gates
(TG$_{1}$, TG$_{2}$) to set the carrier type and density only below their overlap region with the graphene flake.
By controlling the density under the two top gates, various conductance plateaus can be observed in quantum Hall regime. We explain these results by mixing and partitioning of the edge channels at the junctions. Our analysis on these two probe devices indicates that aspect ratio and inhomogeneity play an important role in determining the quantization of the conductance plateaus. The issues we explore here are also important in the use of multiple graphene $p-n$ junctions for metrology \cite{woszczyna_graphene_2011}, use of graphene heterojunctions for collimation \cite{MarcusCollimation} and broad field of metamaterials based on graphene heterostructures \cite{vakil_transformation_2011}.

\begin{figure}
\includegraphics[width=80mm]{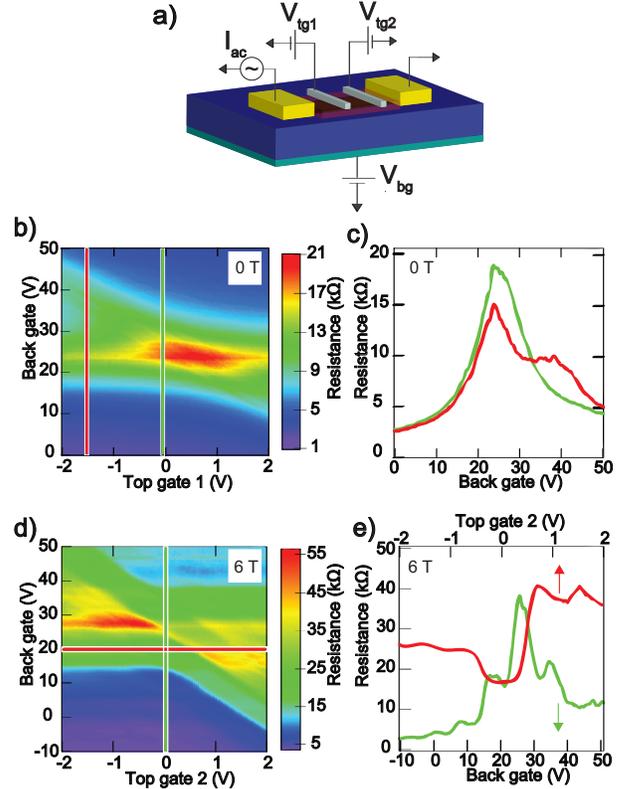}\label{Fig:Fig1}
\caption{(Color online) (a) Schematic diagram of the device. (b) Colorscale plot of two probe resistance as a function of $V_{bg}$ and $V_{tg1}$ at T=1.7~K with $V_{tg2}=$0~V. (c) Line plots of  slices of data shown in (b) at indicated top gate voltages. (d) Colorscale plot of resistance as a function of $V_{bg}$ and $V_{tg1}$ at B=6~T. (e) Line plots of slices of data shown in (d) as indicated by the colored lines.}
\end{figure}

The device fabrication starts with the mechanical exfoliation of graphene flakes from graphite \cite{novoselov_two-dimensional_2005,zhang_experimental_2005} on 300~nm SiO$_2$ grown on degenerately doped silicon substrates.
Using e-beam lithography, source-drain contacts were fabricated by depositing 10~nm/50~nm of Cr/Au. For
the fabrication of top gates, we first spin coat 30~nm of NFC 1400 (JSR Micro) \cite{farmer_utilization_2009} as a
buffer layer followed by the 10~nm of HfO$_2$ using atomic layer deposition to serve as the dielectric.
Following this we fabricate two 500~nm wide top gates with a gap of 2~$\mu$m.
Fig. 1(a) shows a schematic of the device where, $V_{bg}(V_{tg1}, V_{tg2})$ is the back (top) gate bias and $I_{ac}$
($\sim$50~nA) is used to measure the two probe resistance of the device by the AC modulation technique.

To understand the effect of local charge density and type modulation  we start by measuring the resistance with voltages applied
at back gate and at one of the top gates while the other top gate voltage set to 0~V. Fig. 1(b) shows
the colorscale plot of such a measurement. The global maximum in resistance corresponds to the charge neutrality point due to $V_{bg}$ variation. We also see local maxima in resistance (reflected in the angular band on the
colorscale plot) due to the charge neutrality under the top gate. From the resistance variation with back
gate voltage, we measure the carrier mobility to be $\sim$4800~cm$^{2}$/Vs.
The relative capacitive coupling ($\eta$) of the top gate with respect to the back gate is $\sim$7 and can be calculated from
the slope of the two resistance peaks due to the top gate and back gate \cite{huard_transport_2007}. Capacitance calculated by taking into account the dielectric
coefficient of HfO$_{2}$ and NFC \cite{farmer_utilization_2009} gives a $\eta$ of 5.8.
Fig. 1(c) shows line plots from the colorscale plot in Fig. 1(b). The green colored curve in Fig. 1(c) shows the gating effect with the top gate at 0~V hence having minimal contribution to the charge density in the device. In the red curve however there is a clear local maximum at $\sim$40~V which is due to the effect of the top gate which is biased at 1.5~V.

\begin{figure}
\includegraphics[width=80mm]{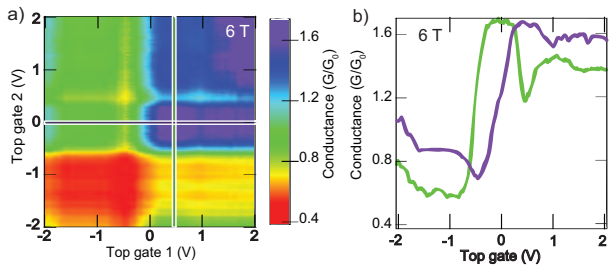}\label{Fig:Fig2}
\caption{(Color online) (a) Measured conductance (in units of $G_0=e^2/h$) of the device with the voltages applied at two top gates while $V_{bg}$=31.1~V corresponding to the $\nu$ =2 at T=1.7~K and B=6~T. (b) Line plots from (a) at the slices indicated by the colored lines on it to show the observed conductance plateaus.}
\end{figure}

Fig. 1(d) shows the colorscale plot of resistance of the device with voltages applied at back gate and at one of the top gates (while other top gate voltage is set to 0~V) in the presence of a 6~T magnetic field applied perpendicular to the plane of graphene. The first thing to be noted is the formation of the quantum Hall plateaus when the top gate is fixed at 0~V.  Fig. 1(e) shows line plots of Fig. 1(d) with the green plot showing the plateaus as a function of the back gate. The plateaus corresponding to $\nu$ of 2 and 6 can be seen. Also seen at V$_{tg}\ensuremath{\neq}$0V are the fractionally quantized conductance plateaus which arise due to the equilibration of the edge channels interacting at the junctions \cite{oezyilmaz_electronic_2007}. The plateaus occurs at slightly higher values than that expected from the exact quantization. This is  possibly due to finite contact resistance, as this is a two probe measurement. Also, finite longitudinal conductivity ($\sigma_{xx}\ensuremath{\neq}0$) in rectangular graphene devices can cause deviation from the ideal quantization \cite{oezyilmaz_electronic_2007,abanin_conformal}. The red plot in Fig. 1(e) shows the plateaus that are not expected for monolayer graphene and arise due to the effect of the top gate.

We next consider the central experiment of this letter which is the interaction of the edge channels induced by two top gates. Fig. 2(a) shows the conductance(G) of the device (in units of  ($G_0=e^2/h$)) as a function of $V_{tg1}$ and $V_{tg2}$ with the back gate ($V_{bg}=31.1$~V) fixing the overall flake at the $\nu$=2 plateau at B=6~T. We observe many fractionally quantized conductance plateaus arising due to the interactions between the edge channels induced below the two top gates mediated via the intermediate graphene lead. Fig. 2(b) shows line plots for slices of data shown in Fig. 2(a) with one of the top gates being varied continuously, while the other top gate and back gate are set at $\nu$=2 plateau.

Another feature of the data shown in Fig. 2(a) is that there are fluctuations in the conductance plateau values that extend over a range of top gate values; seen as horizontal bands at fixed values of $V_{tg2}$. This observation can be understood as resonant reflection due to impurities embedded underneath the top gate \cite{jain_qhe} (we discuss the role of impurities later in detail). Varying
$V_{tg2}$ modifies the chemical potential and that changes the amplitude of scattering from delocalized to localized states.


\begin{figure}
\includegraphics[width=80mm]{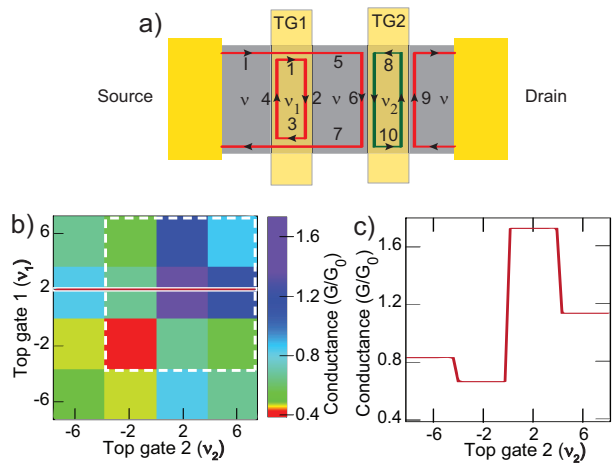}\label{Fig:Fig3}
\caption{(Color online) (a) Schematic of the model employed to calculate the conductance. It shows the various edge channels and the possible types of equilibration at the junctions. (b) Colorscale conductance plot of the results of the calculation as a function of the two top gates while $V_{bg}$ set at $\nu$=2 by taking into account the aspect ratio of the locally gated regions. The dotted line indicates the region in filling factor space probed in the experiment. (c) Line plot showing the slice of colorscale plot in (b) along the marked line.}
\end{figure}

To explain the observed data we employ a simple model as shown in Fig. 3(a). Here the red/green channels represent electrons/holes. This shows just one of the various possible configurations of the regions \cite{abanin_quantized_2007}. We now solve the system of equations framed using current conservation at each of the junctions with the reflection coefficients
of the edge channel currents determined by the type of charge and also the number of channels present. In all there are three configurations of junctions possible. Let $\nu_1$ and $\nu_2$ be the number of edge channels on either side of the junction with $\nu_2$ being below a top gate. The first possibility is when both $\nu_1$ and $\nu_2$ are of the same carrier type but $|\nu_2|<|\nu_1|$ in which case only the channels present in $\nu_2$ are transmitted giving a reflection coefficient $ r_1$ = 1-$\frac{\nu_2}{\nu_1}$ into region 1. The next case is when $|\nu_1|<|\nu_2|$ leading to circulating channels from among $\nu_2$ in region 2. This case is shown in Fig. 3(a) between $\nu$ and $\nu_1$ where $ r_1$ = 1-$\frac{\nu_1}{\nu}$ into the region $TG_{1}$. In our device such circulating channels can be formed even in the region between $TG_{1}$ and $TG_{2}$ for the case when $|\nu_1|<|\nu|$. The remaining possibility is that of $\nu_1$ and $\nu_2$
being of different charge carrier types, i.e. electrons on one side and holes on the other side. This leads to the complete mixing of electron and hole channels at the junction. In Fig. 3(a) this is shown
by $\nu$ and $\nu_2$ giving a reflection coefficient of r$_{2}$= $\frac{\nu_2}{\nu_2+\nu}$
into region below $TG_{2}$. Using current conservation at each interface leads to the following set of equations; $I_1 = I + I_4$, $I_2 = r_1 \times I_1$, $I_3 = I_2 + I_7$, $I_4 = r_1 \times I_3$, $I_5 = I_1 - I_2$, $I_6 = I_5 + I_8$, $I_7 = I_6 - I_{10}$, $I_8 = r_2 \times I_9$,  $I_9 = I_{10}$ and $I_{10} = r_2 \times I_6$,  where $I_i$ represent the current due to channel $i$ as labeled in Fig 3(a). Solving this system of equations gives an effective filling factor $\nu_{eff} = \frac{\nu\nu_1\nu_2}{\nu\nu_1 - \nu\nu_2 + 3\nu_1\nu_2}$. Similarly solving for all the possible configuration of the local filling factors leads to the effective filling factors \cite{Supple}. This however does not match exactly with the experimental data. To account for the deviation we take into account the aspect ratio and incorporate a W/L (device width to length ratio) of 3 and impurity doping in the two top gated regions \cite{abanin_conformal,Supple}. A measure of the inhomogeneity in charge density can be obtained from the width of the Dirac peak as a function of $V_{tg1}$ and $V_{tg2}$ ($\Delta n=C_{tg} \times \Delta V_{tg}$). For the two top gates, $\Delta n$ turns out to be 7.9x10$^{11}$/cm$^{2}$ and 5.4x10$^{11}$/cm$^{2}$. This gives an estimate of the impurity doping in the sample. Using the effective filling factors now obtained we calculate the conductance and plot it in Fig. 3(b). Comparing Fig. 2(b) with the line plot of Fig. 3(b) shown in Fig. 3(c), we see a reasonable match between them. Further refinement of the calculation can be done \cite{Supple}. This calculation is however still unable to explain the diagonal asymmetry and the peaks and valleys in Fig. 2(a) deviating from flat plateaus in conductance as expected from Fig. 3(b). There could be three possible mechanisms
 that can cause such a deviation -- Firstly, depending on the geometry of the flake this quantization is susceptible when aspect ratio deviates from unity \cite{abanin_conformal,abanin_marcus}. Secondly, it can arise due to inhomogeneities present under the locally top gated region, which can affect the exact quantization of the plateaus. The distribution of the impurity below the two top gates can lead to differences in their effect on the conductance quantization, including the oscillations seen in the modulation due to $TG_{2}$ which is absent in the modulation due to $TG_{1}$ \cite{jain_qhe}. This shows that there is an asymmetry in the properties of the regions below the two top gates leading to the asymmetry seen in Fig. 2(a). Further, the $\nu=2$ plateau has been found to be more susceptible to inhomogeneity \cite{ozyilmaz_electronic_2007} under the locally gated region as compared to the other plateaus when the leads are set at the same $\nu$, which is $\nu=2$. Lastly, we speculate that the mismatch in the conductance plateaus from ideal values is possibly due to unintentional dopants leading to multiple uncontrolled $p-n$ junctions \cite{rossi_effective_2009,rossi_signatures_2010}.

\begin{figure}
\includegraphics[width=80mm]{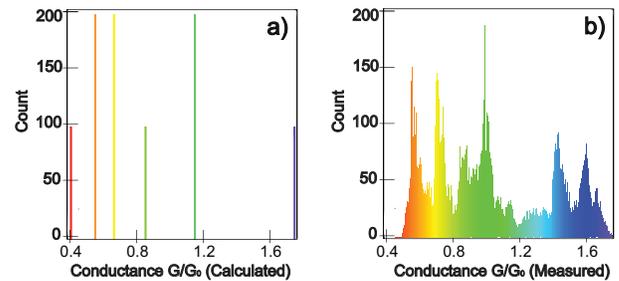}\label{Fig:Fig4}
\caption{(Color online) (a) Histogram of the conductance data plotted in Fig. 3(b). (b) Plots a histogram of the experimentally measured conductance shown in Fig. 2(a).}
\end{figure}

Fig. 4 shows histograms of the conductance plots, both experimental and calculated. It is clear that there is a good correspondence between the two in terms of the overall trend and the expected features. The peaks at the calculated filling factor values can be seen in the histogram from the measured data. The histogram also gives a sense for the mismatch present in the sample shown by the spread of the expected peaks.

In summary, we have studied the effect of two independently locally gated regions on the conductance of graphene in quantum Hall regime.  We have been able to study the equilibration of the channels at multiple junctions giving rise to fractional quantized conductance plateaus and also the critical role that impurity plays. This will help in developing a better understanding of lateral heterostructures for applications like metrology \cite{woszczyna_graphene_2011}.

We acknowledge financial support from the Government of India.

%

\end{document}